# Chiral spin textures creation and dynamics in a rectangular nanostructure

Sateesh Kandukuri[1], Felipe Garcia-Sanchez[2], P K Thiruvikraman[1], V Satya Narayana Murthy*[1],

[1]Department of Physics, BITS Pilani Hyderabad Campus, Medchal District – 500078, Telangana, India

*[1]satyam@hyderabad.bits-pilani.ac.in

[2]Department of Applied Physics, Universidad de Salamanca, 37008 Salamanca, Spain

Abstract

Controlled creation of stable chiral spin textures is required to use them as an energy-efficient information carrier in spintronics. Here we have studied the stable creation of isolated chiral spin texture (skyrmion and antiskyrmion) and its pair through the magnetization reversal of a rectangular nanostructure using spin-polarized currents. An isolated spin texture is created through a negative current pulse. Dynamics of the stable spin texture are explored under external magnetic fields, and the resonant frequencies are calculated. A stable skyrmion pair is created using an asymmetric current pulse, and their interaction is studied using the Thiele equation. The stability of isolated or paired spin texture depends on the DMI strength, spin-polarized current density, and pulse duration. In addition, the stability of the skyrmion pair depends on their initial separation, and a threshold for the separation between skyrmions of 78 nm is observed.

## I. Introduction

Skyrmions are topologically protected chiral vortex spin textures in magnetic systems [1, 2]. In materials with broken inversion symmetry (B20 compounds – MnSi [3] and FeGe [4, 5]) or at the interface of ultrathin heavy metal and ferromagnet (W/ Fe [6], Ir/Fe [7], Pt/Co [8-11] and Pd/Co [12]), spin-orbit coupling induces a sizeable asymmetric Dzyaloshinskii – Moria interaction (DMI) [13-15], which is responsible for the formation of magnetic skyrmion spin textures. More recently, an exotic relative spin texture antiskyrmion has been observed in $D_{2d}$ symmetry Heusler alloys [16, 17] and $C_{2v}$ symmetry thin films [18, 19]. Depending on the crystal symmetry, a magnetic system can host a variety of spin textures [20]. These chiral spin textures are characterized by an invariant integer called topological charge $Q$ [1, 2 and 21], which can be calculated from $Q = \frac{1}{4\pi} \int \boldsymbol{m}.(\partial_x \boldsymbol{m} \times \partial_y \boldsymbol{m}) \, dxdy$ and for a structure like a skyrmion in an ultrathin film can have values $Q = \pm 1$. That charge counts how many times normalized magnetization $\boldsymbol{m}$ wraps the unit sphere. The boundaries of the skyrmion and antiskyrmions have distinct chirality, and they have opposite topological charges under the same core magnetization [1, 22]

Magnetic skyrmions have drawn attention in spintronics due to their small stable size in the nanometer range and high mobility for low current densities ($10^6 \, A/m^2$) [23]. To use it as an energy-efficient information carrier [24 - 27], controlled creation and manipulation of chiral spin textures are required. A recent review article discusses the various methods of magnetic skyrmion creation including external magnetic fields, spin-polarized currents, local electric fields, and lasers [28]. Several papers [29 - 32] have reported the chiral spin textures creation using spin transfer torques exerted by the spin-polarized currents [33 - 37]. It is observed that the spin texture is created mainly by reversing the magnetization at the desired location using nano-contacts. In our previous work, we reported the creation of chiral spin textures by flipping the entire confined square nanostructure magnetization [38]. The skyrmion spin texture showed a breathing mode in the relaxation. Those modes were studied by Joo Von Kim et al. in a circular nano dot structure [39]. Previously, D Capic et al. reported that a skyrmion pair with a finite separation in a square thin film is relaxed through the gyrotropic motion [40]. Our present study observes similar behaviour, which is discussed later.

Here we present the creation of chiral spin textures in a rectangular Co/Pd nanostructure [12] through the magnetization reversal using spin-polarized currents. An unstable antiskyrmion evolution and breathing mode of the skyrmion during relaxation is observed while creating a skyrmion. Lorenzo Camosi et al. reported that thin film systems with C2v symmetry induce anisotropic DMI [18]. We used a DMI with an opposite sign along the two perpendiculars in-plane directions to study the effect of spin texture formation, and an anisotropic DMI favoured the stable antiskyrmion formation. Further, the dynamic modes of isolated skyrmion and antiskyrmion spin textures are explored with external magnetic fields. Finally, we showed the creation of a chiral spin texture pair with an asymmetric current pulse and explained the interaction between the skyrmion pair using the Thiele equation [41].

This paper is organized as follows - Sec. II, a description of the model and simulations are explained. The skyrmion and antiskyrmion creation through spin-polarized current is explained in Sec. III. The dynamics of the isolated spin texture under external magnetic fields are described in Sec. IV. Chiral spin texture pair creation and interaction are analyzed in Sec. V. Finally, some concluding remarks are given in Sec. VI.

**II. Model and Simulations**

Chiral spin texture creation and dynamics are studied on a rectangular nanopillar structure using the micromagnetic simulation software mumax3 [42, 43]. The nanostructure comprises a thick reference layer at the bottom and a thin free layer at the top, separated by a spacer layer (see Fig.1). The magnetization of the free ($m$) and reference layer ($m_p$) are defined perpendicular to the plane of the nanostructure. The free layer's magnetization is manipulated through spin-polarized currents for the chiral spin textures creation. The energy minimization dynamics of the free layer are described by Landau-Lifshitz-Gilbert's equation with Slonczweski spin transfer torque (LLGS) [29, 32, 44],

$$\frac{d\boldsymbol{m}}{dt} = -\frac{\gamma_0}{1+\alpha^2}\left(\boldsymbol{m}\times\boldsymbol{H}_{eff} + \alpha(\boldsymbol{m}\times(\boldsymbol{m}\times\boldsymbol{H}_{eff}))\right) + \boldsymbol{\mathcal{T}}_{STT}$$

where $\boldsymbol{m} = \vec{M}/M_s$ is a normalized unit vector of magnetization and $M_s$ is the saturation magnetization. $\gamma_0$ is the gyromagnetic constant, and the first part describes the precession of magnetization under the effective field $\mathbf{H}_{eff}$. It is a variational derivative of the total micromagnetic energy function $U$ concerning $\boldsymbol{m}$,

$$\mathbf{H}_{eff} = -\frac{1}{\mu_0 M_s}\frac{\delta U/V}{\delta \boldsymbol{m}}$$

Here the total magnetic energy $U$ considers exchange interaction, dipole-dipole interaction, magnetocrystalline anisotropy energy and Zeeman energy. It also includes the interaction from interfacial DMI [45],

$$U_{\text{DMI}} = l\iint dxdy\,\left(D_x(m_z\frac{\partial m_x}{\partial x} - m_x\frac{\partial m_z}{\partial x}) + D_y(m_z\frac{\partial m_y}{\partial y} - m_y\frac{\partial m_z}{\partial y})\right)$$

where $l$ is the thickness of the free layer, $D_x$ and $D_y$ are in-plane components of the DMI. The DMI couplings are known as Lifshitz invariants expressed in antisymmetric differential form [46]. For $D_x = D_y$, the DMI is isotropic, and it favours the stabilization of skyrmions. On the other hand, anisotropic DMI $(D_x, -D_y)$, favours the formation of antiskyrmions.

The second part of the LLGS equation represents the magnetization damping and is characterized by the Gilbert damping parameter $\alpha$. The third part is the Slonczweski spin transfer torque ($\boldsymbol{\mathcal{T}}_{STT}$) that arises from the spin-polarized currents. It is defined as [42],

$$\boldsymbol{\mathcal{T}}_{STT} = \beta\frac{\varepsilon - \alpha\varepsilon'}{1+\alpha^2}\left(\boldsymbol{m}\times(\boldsymbol{m}_p\times\boldsymbol{m})\right) - \beta\frac{\varepsilon' - \alpha\varepsilon}{1+\alpha^2}(\boldsymbol{m}\times\boldsymbol{m}_p)$$

$$\beta = \frac{\gamma \hbar J_z}{M_s e l}, \quad \varepsilon = \frac{P}{2}$$

where $J_z$ is the current density along the z direction, P spin polarization, $\varepsilon'$ is the secondary spin transfer torque parameter, and $\boldsymbol{m_p}$ is the magnetization of the reference layer. Simulations are performed on free layer dimensions of 250 nm x 150 nm x 3 nm with discretized cell sizes of 1nm x 1nm x 1nm. The material parameters are considered from a symmetric Pd/Co/Pd multilayer system [12], which possesses an interfacial DMI required to form chiral spin textures. We considered isotropic DMI for creating skyrmions and anisotropic DMI for antiskyrmion with an exchange constant of A = 15 pJ/m, a saturation magnetization of $M_s$ = 280 kA/m, a uniaxial anisotropy of $K_u$ = 0.06 MJ/m$^3$, and a Gilbert damping constant of $\alpha$ = 0.1. The following sections are devoted to the study of chiral spin texture formation and its dynamics using the current densities range of -0.1 x 10$^{12}$ to -5.0 x 10$^{12}$ A/m$^2$ and different pulse widths (0.5 ns for isolated skyrmions and 30 – 150 ps for skyrmion pair) for DMI magnitudes from 0.30 to 0.50 mJ/m$^2$. For isolated chiral spin texture creation, the spin-polarized current is applied in a negative z-direction $(-J_z)$, and for the spin texture pair creation, asymmetric pulses are applied. We used the same spin polarization of P = 0.4 and $\varepsilon' = 0$ in all the cases.

### III. Creation of an Isolated Chiral Spin Texture

**(a) Skyrmion**

Initially, the free layer is uniformly magnetized ($\hat{z}$) and then relaxed to obtain the equilibrium state in the presence of isotropic DMI ($D_i$). Then, a spin-polarized current density is applied for 0.5 ns in this study. Depending on the current value and the DMI constant, we can find different final states. The current density ($J_z$) up to -0.3 x 10$^{12}$ A/m$^2$ didn't induce the magnetization reversal for any of the DMI values. For $J_z$ = -0.4 x 10$^{12}$ A/m$^2$, $D_i$ = 0.30 and 0.35 mJ/m$^2$, magnetization reversal started across the opposite corners and moved along the edges to the other corners. Due to the short current pulse width (0.5 ns), this reversal is annihilated later. For $D_i$ = 0.40 to 0.50 mJ/m², spin reversal merging from opposite corners seems to create an antiskyrmion, but it collapsed immediately.

The spin states and topological charge variation for $J_z$ = -0.5 x 10¹² A/m² are shown in Fig. 1a and 1b, respectively. For all DMIs, an unstable antiskyrmion evolved from spin reversal merging at the end of the pulse. The antiskyrmion annihilated at 2.24 ns for 0.30 mJ/m² DMI. It is eradicated through expansion for 0.35 and 0.40 mJ/m² DMI and collapsed immediately for 0.45 and 0.50 mJ/m² DMI. Figures 2a and 2b show, respectively, the spin states and the corresponding topological charge variation for $J_z$ = -0.6 x 10¹² A/m². In the case of $D_i$ = 0.30 mJ/m², spin reversal formed a stripe-like domain, which is ripped off into a skyrmion and two antiskyrmions. At first, the skyrmion, followed by the antiskyrmions, are eradicated. A skyrmion is formed within the pulse for $D_i$ = 0.35 - 0.50 mJ/m2. Moreover, it is stabilized through breathing mode in the relaxation process for $D_i$ = 0.35 – 0.45 mJ/m². That is confirmed by the topological charge at 5ns in Fig. 2b. For 0.50 mJ/m² DMI, the skyrmion is annihilated by expanding its core. Figure 2c explains the core size variation of a stable skyrmion, and clearly, the skyrmion size increases as the DMI strength increases, as expected due to the role of DMI.

The phase diagram of skyrmion formation for the isotropic DMIs from 0.30 to 0.50 mJ/m² and different spin-polarized current densities applied for 0.5 ns is shown in Fig. 3a. The current densities $J_z \geq$ -0.7 x 10¹² A/m² created a skyrmion for all the DMIs. The skyrmion is destroyed within the current pulse width for $D_i$ = 0.30 mJ/m2 and $J_z \geq$ -0.7 x 10¹² A/m2. A stable skyrmion is observed for $D_i$ = 0.35 mJ/m² at $J_z$ = -0.8 and -0.9 x 10¹² A/m² and for $D_i$ = 0.40 mJ/m² from $J_z$ = -0.8 to -1.1 x 10¹² A/m². Above these current densities, the skyrmion is annihilated within the current pulse. In the case of $D_i$ = 0.45 and 0.50 mJ/m², the skyrmion is unstable and eradicated through the expansion of the core during the relaxation.

**(b). Antiskyrmion**

We implemented a modified mumax3 code for anisotropic DMI favouring antiskyrmion formation. The in-plane DMI component magnitudes are considered equally $(\|D_x\| = \|D_y\| = D_a)$ to compare the dynamics of skyrmions and antiskyrmions. Isolated antiskyrmion formation for different current densities is similar to the skyrmion formation. Notedly, the unstable skyrmion formation is observed for $J_z$ = -0.5 x 10¹² A/m² and for all the anisotropic DMIs is shown in supplementary Fig. S1 (a) (The corresponding topological charge variation is shown in supplementary Fig. S1 (b)). For $D_a$ = 0.30 mJ/m² and $J_z$ = -0.6 x 10¹² A/m², a stripe-like domain evolved and ripped out into an antiskyrmion and two skyrmions (Supplementary Fig. S2 (a) and (b)). Under the same material parameters and with an equal magnitude of isotropic and anisotropic forms of DMI, the core size of the antiskyrmion is slightly larger than the skyrmion size (Supplementary Fig. S2 (c)). The phase diagrams of the skyrmion (Fig. 3a) and antiskyrmion (Fig. 3b) show that the formation of a particular chiral spin texture depends on the nature of the DMI. In the regions SR1 and ASR1, the spin texture nucleation is not observed. Merging of magnetization reversal from the corners created an unstable skyrmion and antiskyrmion phase in SR2 and ASR2 regions. Even though skyrmion and antiskyrmion nucleation happen in the respective regions of SR3 and ASR3, they are annihilated by shrinking their size within the current pulse time. In the regions SR4 and ASR4, stable skyrmion and antiskyrmion nucleation is observed, respectively.

**IV. Dynamics under perpendicular sinusoidal fields**

The spin-polarized current nucleated chiral spin textures are stabilized through the breathing mode. Due to that, in this section we explore the excitation dynamics under ac magnetic fields with isotropic DMI of 0.35 mJm$^{-2}$ for skyrmion and anisotropic DMI of 0.35 mJm$^{-2}$ for antiskyrmion. Initially, the stability of the spin textures is tested by applying a static uniform magnetic field perpendicular to the system. As the spin direction at the core centre is upwards, the core size is increased for the positive field and decreased for the opposite field. The skyrmion is stable for the constant field range of -40 to 12 mT, and the antiskyrmion is stable between -52 to 10 mT (see Fig. 4). Above these high positive fields, the spin texture is unstable because the system width is less than the length of the chiral structure. Under the static perpendicular magnetic field ($B_Z$), the dynamic modes of the spin textures are computed from their transient response to an applied field in the form $h_z(t) = h_0 \sin(2\pi f_c t)/(2\pi f_c t)$. With 0.1 mT excitation field's amplitude ($h_0$) and 50 GHz cut-off frequency ($f_c$), the dynamics are evaluated for 20 ns. Fig. 5 shows the power spectral density (PSD) calculated for excitation modes by employing the Fast Fourier Transform (FFT) on transient dynamic magnetization. The PSD plot displays four discrete modes ($f_1$ - $f_4$), whose frequencies exhibit a linear dependence on the static field.

Further, the spatial dynamic behaviour of the spin texture is studied using the sinusoidal magnetic field $h_z(t) = h_0 \sin(2\pi f_r t)$ with the resonant mode frequencies obtained earlier. After the field is applied for 20 ns, magnetic states are saved at time intervals of $1/20 f_r$ over ten oscillation periods. The spatial fluctuations of the magnetization are calculated using $\delta m(t) = m(0) - m(t)$, where $m(0)$ represents the initial equilibrium state. Fig. 6a shows the spatial fluctuations of three magnetization components at one instant for the four resonant mode frequencies under zero static magnetic field of the skyrmion spin texture. Fig. 6b represents the $m_z$ component fluctuation ($\delta m_z(t)$) along the x-axis, passing through the centre of the nanostructure for one period of oscillation. The skyrmion and antiskyrmion spin textures show similar dynamic responses to the sinusoidal excitation fields. For mode one ($f_1$), the spatial fluctuation of the magnetization in Fig. 6a and also from the plot of $\delta m_z(x, t)$ in Fig. 6b show that the dynamic response is predominately from the spin texture. As the core magnetization expands and contracts, one can conclude that mode one is the breathing mode of the spin texture. The other three modes ($f_2 - f_4$) are hybrid due to the nanostructure edges excitation with the spin texture's breathing mode, as previously reported by Kim et al [39]. Mainly, the short sides of the system in mode two ($f_2$), the corners in mode three ($f_3$) and the system edge centres in mode four ($f_4$) are excited along with the breathing mode of the spin texture. We observed similar dynamic excitations for the corresponding modes of the spin texture under the stable static field range ($B_z$).

**V. Chiral Spin Texture Pair Creation**

In the isolated chiral spin texture creation, the magnetization reversal is initiated from the corners of the system and it happens earlier for a larger value of DMI. It is due to spins tilting more strongly at the nanostructure boundaries for the higher DMI (see Fig. 7a). To see how the initial spin state configuration affects the chiral spin texture creation, we applied a short positive current pulse ($+\hat{z}$) to saturate the state and then we applied a negative pulse ($-\hat{z}$). We observed chiral spin texture pair formation for the negative current densities $\geq 2.0 \times 10^{12}$ A/m$^2$ and the pulse widths $< 150$ ps. Spin reversal induced a stripe-like domain, which is ripped off into an antiskyrmion and two skyrmions for isotropic DMI. The strip is ripped off into a skyrmion and two antiskyrmions for anisotropic DMI. In the case of isotropic DMI, antiskyrmion annihilated immediately. The skyrmion pair stabilized in some cases during the relaxation (see Fig. 7b) and the behaviour is opposite to the anisotropic DMI. Fig.8 shows the skyrmion pair dynamics for 10 and 50 ps positive pulse ($t_p$) and 82 ps negative pulse ($t_n$) with a current density of 3.0 x $10^{12}$ A/m$^2$ and $D_i$ = 0.35 mJ/m$^2$. Once the current pulse is off, due to repulsive interaction, the skyrmion pair moved away from each other in a gyrotropic path (see Fig. 8a and 8c) by showing a breathing mode of nearly 5 ns (see Fig. 8b and 8d). The skyrmion pair annihilated at the edges for a 10 ps positive pulse (tp), and for 50 ps it is stabilized. We assign this behaviour to the repulsive interaction, which depends on the initial separation between the skyrmion pair ($r_i$) – 76 nm for 10 ps and 87 nm for 50 ps.

As similar behaviour in the creation of skyrmion pair and antiskyrmion pair are observed, further studies are carried out for fixed isotropic DMI of 0.35 mJ/m$^2$ for the positive current densities from 1.0 to 5.0 x $10^{12}$ A/m$^2$ (+$J_z$) applied for 10 - 90 ps ($t_p$), and negative current densities from 2.0 to 4.0 x $10^{12}$ A/m$^2$ (-$J_z$) applied for 40 – 150 ps ($t_n$). Figs. 9a, b, and c show the minimum and maximum negative pulse time ($t_n$) dependence on the positive current density (+$J_z$) and its pulse ($t_p$) for the skyrmion pair formation respectively for 2.0, 3.0 and 4.0 (x $10^{12}$) A/m$^2$ negative current densities. The minimum negative current pulse time (min. $t_n$: hollow points in Figs. 9a, 9b and 9c) required for observing the skyrmion pair decreases with increasing positive current density (+$J_z$) and negative current density (-$J_z$). The maximum negative current pulse time (max. $t_n$: solid points in Figs. 9a, 9b and 9c) is also reduced, except for the case of negative current density 2 x $10^{12}$ A/m$^2$ (see Fig. 9a); this could be due to the strength of the spin transfer torque increasing with the current density. The positive current density (+$J_z$) applied above 10 ps ($t_p$) seems to saturate the free 'layer's initial magnetization state. The minimum and maximum negative current pulse width ($t_n$) saturates for the $t_p$ > 10 ps as positive and negative current densities increase. The initial separation of the skyrmion pair increased with positive current density and pulse, as shown in Figs. 9d, e, and f. We observed a minimum separation of 78 nm between the skyrmions below which the skyrmion pair got annihilated at the edges. The above results were carried out for a zero delay between the positive and negative current pulse, and, even when it was introduced a delay of 10 ps a stable pair of spin textures were created.

**Skyrmion – Skyrmion Interaction**

To understand the interaction between the skyrmion pair more clearly, we used the Thiele approach to evaluate the force of repulsion exerted by the skyrmions on each other. The position of the two skyrmions is defined as $r_1 = (x_1, y_1, 0)$ and $r_2 = (x_2, y_2, 0)$ and their speeds $v_1 = (v_{x,1}, v_{y,1}, 0)$ and $v_2 = (v_{x,2}, v_{y,2}, 0)$. The force can be supposed to be radial and depend only on the distance between the skyrmions $r = |r_1 - r_2|$, like a central force. Then, the force will have opposite directions on each skyrmion and will have the form $F_1 = f(r)\hat{r}$ and $F_2 = -f(r)\hat{r}$, where $\hat{r} = (r_1 - r_2)/r$ is the unit vector pointing from skyrmion-1 to skyrmion-2. One can establish a Thiele equation for each skyrmion.

$$\text{In general,} \quad G \times v - \alpha \widetilde{\mathcal{D}} v = F$$

where $G = \frac{-4\pi Q M_s l}{\gamma}\hat{k}$ is the gyrotropic vector and $\widetilde{\mathcal{D}} \equiv D_{ii}$ is the diagonal component of the dissipative tensor. $D_{ij} = \frac{M_s}{\gamma} \int \frac{\partial m}{\partial x_i} \frac{\partial m}{\partial x_j} dV$ with $x_i$ representing the cartesian coordinates, we always have $D_{xx} = D_{yy} = D$ and $D_{xy} = D_{yx} = 0$ for the skyrmion profile. The Thiele equation for the coupled skyrmions can be rewritten as $G \times \Delta v - \alpha D \Delta v = 2F$. Where $\Delta v = (\Delta v_x, \Delta v_y, 0)$ is the difference in velocities between two skyrmions. We used an FFT filter to smoothen the noisy variation of the skyrmions' position, and from that, we obtained the velocities. From the fitting of the velocities, one can obtain the value of the force $f(r)$.

For the skyrmion pair dynamics explained in Fig.8, the $\Delta v$ and $f(r)$ variations are shown in Fig.10. As the skyrmions in the pair are equal, the force is repulsive and increases up to 1 ns for both unstable and stable creation. The dynamics induced by the repulsion force are heavily damped for the unstable pair and lightly damped for the stable pair. In the case of the unstable pair, due to less initial separation (76 nm), the force of repulsion is larger, leading to the annihilation of the skyrmion pair. It is clear from Fig. 11 that as the initial separation increases, the interaction force is reduced, and above the threshold separation of 78 nm, the skyrmion pair is stabilized.

**VI Conclusions**

We showed the creation of stable chiral spin textures through the magnetization reversal using spin-polarized currents. The magnetic system with an isotropic DMI helps to stabilize the skyrmions, and an anisotropic DMI stabilizes antiskyrmions. Stable isolated spin texture formation mainly depends on the spin-polarized current density and pulse time. During the current pulse, the spin transfer torque compresses the spin texture; once the pulse is off, the spin texture shows a breathing mode up to 5 ns and stabilizes after. As the size of the stable spin texture increases with increasing DMI strength, the isolated chiral spin texture dynamics are explored under ac magnetic fields for DMI 0.35 mJ/m$^2$. The dynamic response is similar for skyrmion and antiskyrmion; a primary breathing mode at 1.0 GHz and three hybrid modes at 3.8, 5.2 and 7.0 GHz under zero static field are observed. The resonant mode frequencies($f_1$-$f_4$) exhibited a linear dependence on the static field, and the spatial dynamics response to the corresponding mode frequencies is similar to the zero static field case. In the isolated spin texture creation, it is observed that the spin reversal is earlier for higher DMI due to stronger tilts at the corner edges of the nanostructure. To see the effect of initial configuration dependence on the spin texture formation, we applied a positive current pulse to saturate the state and then reversed the pulse; a spin texture pair evolved from the stripe-like domain. As the skyrmion and antiskyrmion pair formations are similar, we focused mainly on the stable skyrmion pair formation for the current densities and their pulse widths. Finally, the interaction dynamics of the skyrmion pair are studied with the Thiele equation, and it is observed that beyond a critical separation of 78 nm, the skyrmion pair becomes stable.

**References**

1. N. Nagaosa, Y. Tokura, Nat. Nanotechnol. **8**, 899–911 (2013).

2. A. Fert, N. Reyren, and V. Cros, Nat. Rev. Mater. **2**, 17031 (2017).

3. S. Mühlbauer, B. Binz, F. Jonietz, C. Pfleiderer, A. Rosch, A. Neubauer, R. Georgii, and P. Böni, Science **323**, 915 (2009).

4. X. Z. Yu, N. Kanazawa, Y. Onose, K. Kimoto, W. Z. Zhang, S. Ishiwata, Y. Matsui, and Y. Tokura, Nature Mater. **10**, 106 (2011).

5. S. X. Huang and C. L. Chien, Phys. Rev. Lett. **108**, 267201 (2012).

6. M. Heide, G. Bihlmayer, and S. Blügel, Phys. Rev. B **78,** 140403(R) (2008).

7. S. Heinze, K. von Bergmann, M. Menzel, J. Brede, A. Kubetzka, R. Wiesendanger, G. Bihlmayer, and S. Blügel, Nat. Phys. **7**, 713 (2011).

8. B. Dupé, M. Hoffmann, C. Paillard, and S. Heinze, Nat. Commun. **5**, 4030 (2014).

9. A. Hrabec, N. A. Porter, A. Wells, M. J. Benitez, G. Burnell, S. McVitie, D. McGrouther, T. A. Moore, and C. H. Marrows, Phys. Rev. B **90**, 020402(R) (2014).

10. Kai Di, Vanessa Li Zhang, Hock Siah Lim, Ser Choon Ng, Meng Hau Kuok, Jiawei Yu, Jungbum Yoon, Xuepeng Qiu, and Hyunsoo Yang, Phys. Rev. Lett. 114, 047201 (2015).

11. M. J. Benitez, A. Hrabec, A. P. Mihai, T. A. Moore, G. Burnell, D. McGrouther, C. H. Marrows, and S. McVitie, Nat. Commun. **6** 8957 (2015).

12. J. Brandão, D. A. Dugato, R. L. Seeger, J. C. Denardin, T. J. A. Mori, and J. C. Cezar, Sci. Rep. **9**, 4144 (2019).

13. I. Dzyaloshinsky, J. Phys. Chem. Solids **4**, 241 (1958).

14. T. Moriya, Phys. Rev. **120**, 91 (1960).

15. T. Moriya, Phys. Rev. Lett. **4**, 228 (1960).

16. A. K. Nayak, V. Kumar, T. Ma, P. Werner, E. Pippel, R. Sahoo, F. Damay, U. K. Rößler, C. Felser, and S. S. Parkin, Nature **548**, 561 (2017).

17. Daniil A. Kitchaev and Anton Van der Ve, Phys. Rev. Mater. **5**, 124408 (2021).

18. L. Camosi, S. Rohart, O. Fruchart, S. Pizzini, M. Belmeguenai, Y. Roussigné, A. Stashkevich, S. M. Cherif, L. Ranno, M. de Santis, and J. Vogel, Phys. Rev. B **95**, 214422 (2017).

19. M. Hoffmann, B. Zimmermann, G.P. Müller, D. Schürhoff, N. S. Kiselev, C. Melcher, and S. Blügel, Nat. Commun. **8**, 308 (2017).

20. K. Everschor-Sitte, J. Masell, R. M. Reeve, and M. Kläui, J. Appl. Phys. **124**, 240901 (2018).

21. R. Rajaraman, *Solitons and Instantons* (Elsevier, 1987).

22. W. Koshibae and N. Nagaosa, Nat. Commun. **7**:10542 (2016).

23. F. Jonietz, S. Mühlbauer, C. Pfleiderer, A. Neubauer, W. Münzer, A. Bauer, T. Adams, R. Georgii, P. Böni, and R. A. Duine *et al.*, Science **330**, 1648 (2010).

24. A. Fert, V. Cros, and J. Sampaio, Nat. Nanotechnol. **8**, 152 (2013).

25. S. Parkin and S. H. Yang, Nature Nanotech **10**, 195–198 (2015).

26. X. Zhang, W. Cai, X. Zhang, Z. Wang, Z. Li, Y. Zhang, K. Cao, N. Lei, W. Kang, Y. Zhang, H. Yu, Y. Zhou, and W. Zhao, ACS Appl. Mater. Interfaces **10**, 16887 (2018).

27. N. E. Penthorn, X. Hao, Z. Wang, Y. Huai, and H. W. Jiang, Phys. Rev. Lett. **122**, 257201 (2019).

28. X. Zhang, Y. Zhou, K. M. Song, T. E. Park, J. Xia, M. Ezawa, X. Liu, W. Zhao, G. Zhao, and Seonghoon Woo, J. Phys. Condens. Matter **32** 143001 (2020).

29. Y. Tchoe and J. H. Han, Phys. Rev. B **85**, 174416 (2012).

30. J. Sampaio, V. Cros, S. Rohart, A. Thiaville, and A. Fert, Nat. Nanotechnol. **8**, 839 (2013).

31. N. Romming, C. Hanneken, M. Menzel, J. E. Bickel, B. Wolter, K. von Bergmann, A. Kubetzka, and R. Wiesendanger, Science **341**, 636 (2013).

32. H. Y. Yuan and X. R. Wang Sci. Rep. **6** 22638 (2016).

33. J. Slonczewski, J. Magn. Magn. Mater. **159**, L1 (1996).

34. L. Berger, Phys. Rev. B **54**, 9353 (1996).

35. J. C. Slonczewski, J. Magn. Magn. Mater. **247,** 324 (2002).

36. J. Xiao, A. Zangwill, and M. D. Stiles, Phys. Rev. B **70**, 172405 (2004).

37. A. Brataas, A. Kent, and H. Ohno, Nature Mater **11**, 372–381 (2012).

38. K. Sateesh, V.S.N. Murthy, and P.K. Thiruvikraman, Sci. Rep. **11**, 18945 (2021).

39. J. V. Kim, F. Garcia-Sanchez, J. Sampaio, C. Moreau-Luchaire, V. Cros, and A. Fert, Phys. Rev. B **90**, 064410 (2014).

40. D. Capic, D. A. Garanin and E. M. Chudnovsky J. Phys. Condens. Matter **32** 415803 (2020).

41. A. Thiele, Phys. Rev. Lett. **30**, 230 (1973).

42. A. Vansteenkiste, J. Leliaert, M. Dvornik, M. Helsen, F. Garcia-Sanchez, and B. Van Waeyenberge, AIP Adv. **4**, 107133 (2014).

43. J. Mulkers, B. Van Waeyenberge, and M. V. Milošević, Phys. Rev. B **95**, 144401 (2017).

44. F. Garcia-Sanchez, J. Sampaio, N. Reyren, V. Cros, and J. V. Kim, New J. Phys. **18**, 075011 (2016).

45. C. Moreau-Luchaire, C. Moutafis, N. Reyren, J. Sampaio, C. A. F. Vaz, N. Van Horne, K. Bouzehouane, K. Garcia, C. Deranlot, and P. Warnicke et al., Nat. Nanotechnol. 11, 444 (2016).

46. A. N. Bogdanov and U. K. Rößler, Phys. Rev. Lett. **87**, 037203 (2001).

**Acknowledgements**

We acknowledge the BITS Pilani Hyderabad Campus for providing the Sharanga high-performance computational facility to carry out the above work.

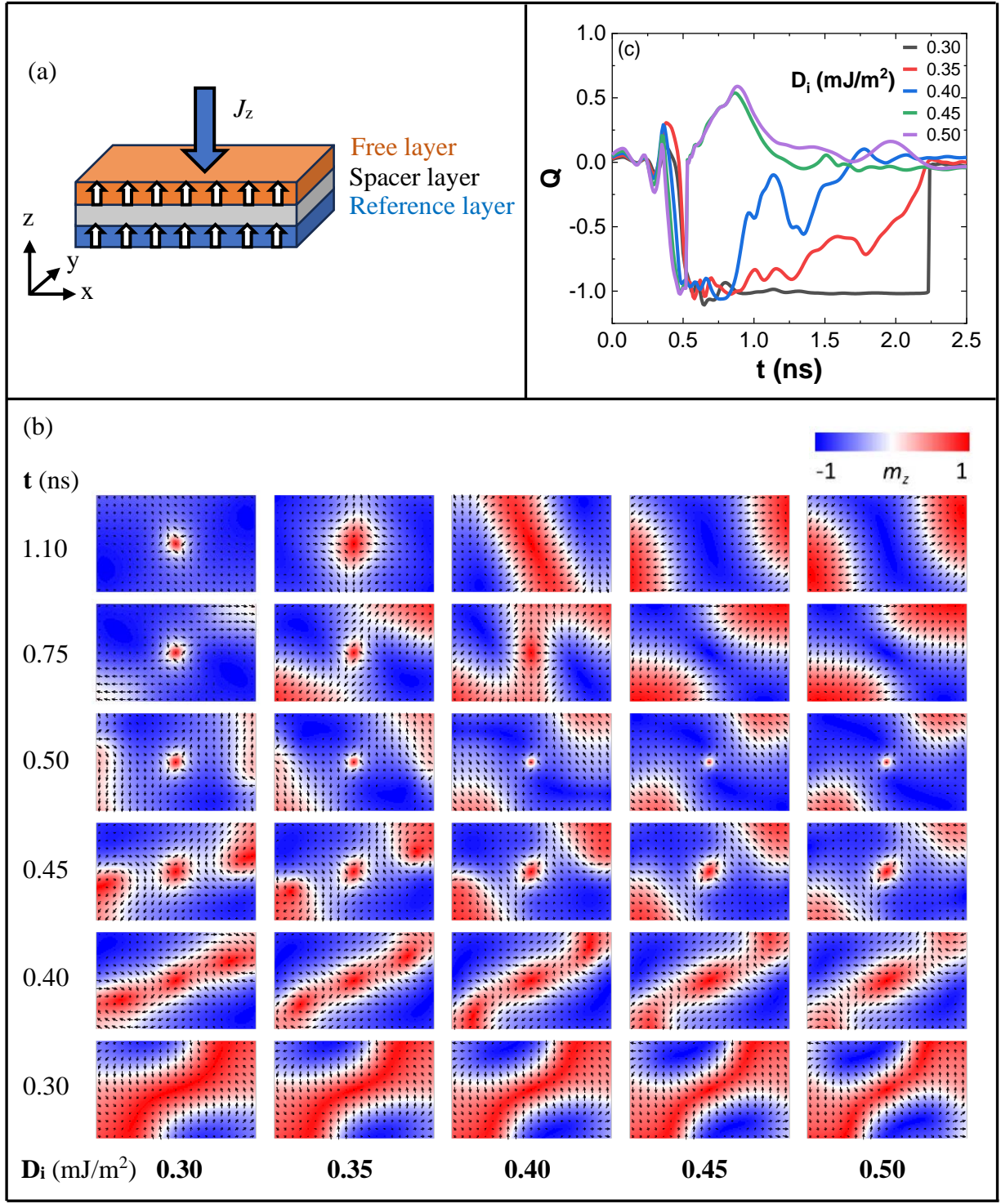

**Fig. 1:** (a) Rectangular nanostructure with spin-polarized current direction. The arrows inside the geometry indicate the initial magnetization directions, and the current flows along the -z direction. (b) and (c) represents the spin states and corresponding topological charge variation for $J_z = -0.5 \times 10^{12}$ A/m$^2$. The spin states at different instants for $D_i = 0.30 - 0.50$ mJ//m$^2$ show the evolution of antiskyrmion. The topological charge $Q = -1$ for the corresponding spin state indicates that the spin texture is an antiskyrmion.

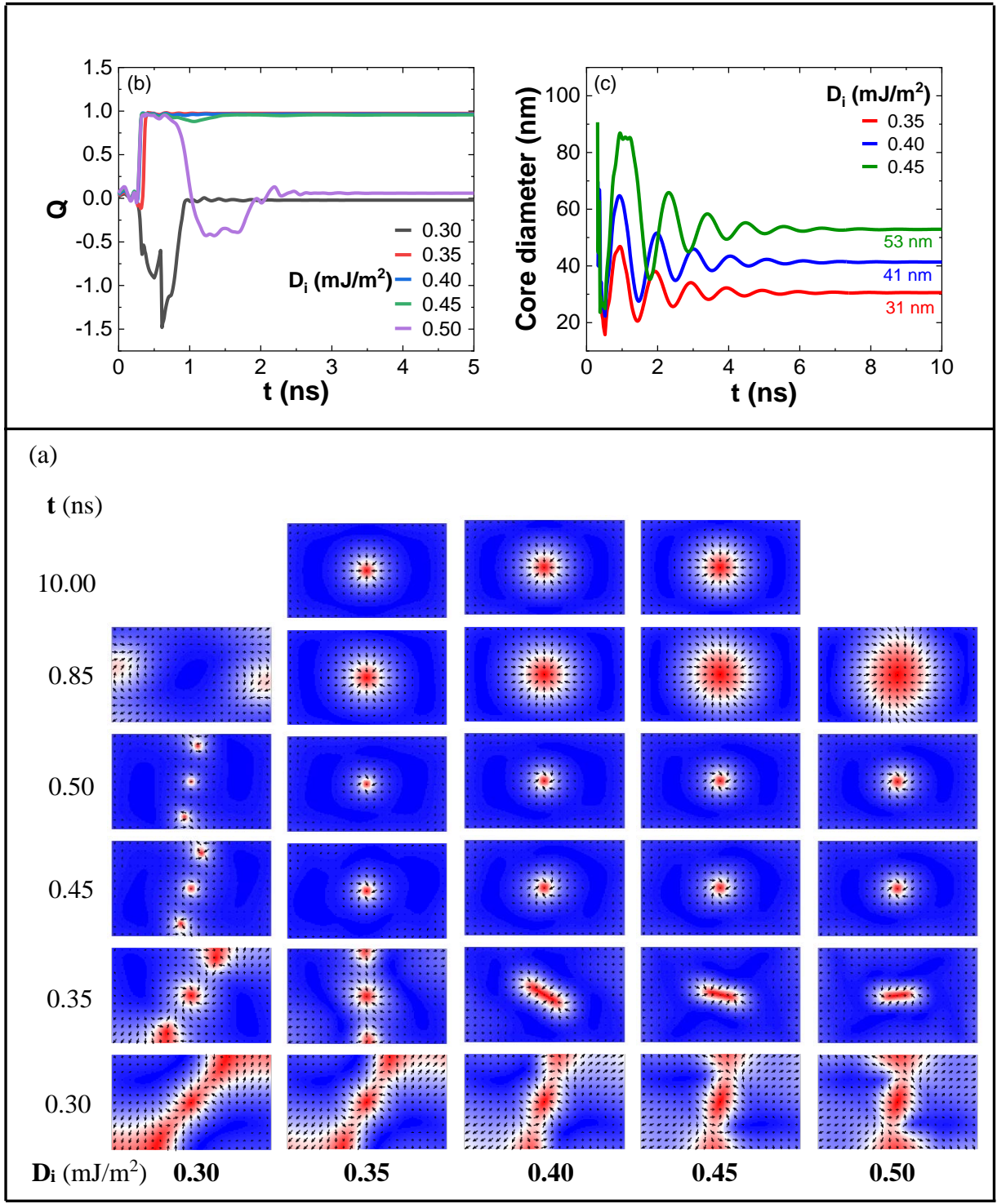

**Fig. 2:** Spin states, topological charge Q and the skyrmion core diameter variation for $J_z = -0.6 \times 10^{12}$ A/m$^2$. (a) Evolution of stripe-like domain ripped off into a skyrmion and two incomplete antiskyrmions for $D_i = 0.30$ and $0.35$ mJ/m$^2$. Skyrmion is stabilized for the $D_i = 0.35 - 0.45$ mJ//m$^2$. (b) The topological charge Q = 1 at 10 ns indicates the stable skyrmion. (c) The core diameter variation upto ~ 5 ns indicates the breathing mode, and the skyrmion size increases with DMI strength.

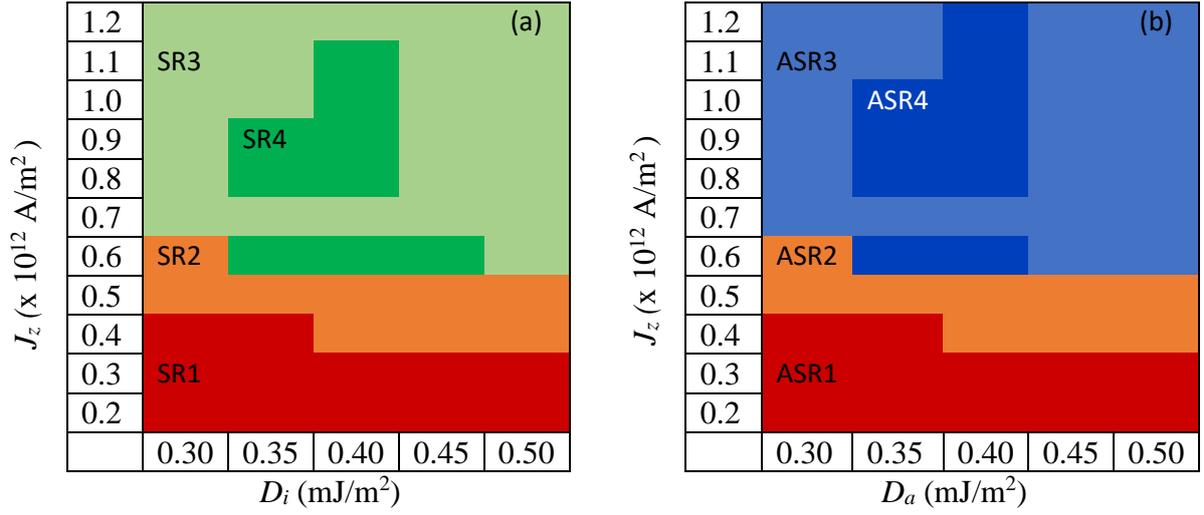

**Fig. 3:** The phase diagram of the skyrmion and antiskyrmion formation.
(a) Skyrmion: SR1 is the region of no-skyrmion, SR2 is the region of skyrmion and antiskyrmion evolution, SR3 and SR4 are the unstable and stable regions of skyrmion formation. (b) Antiskyrmion: ASR1 is the region of no-antiskyrmion, ASR2 is the region of skyrmion and antiskyrmion evolution, ASR3 and ASR4 are the unstable and stable regions of antiskyrmion formation.

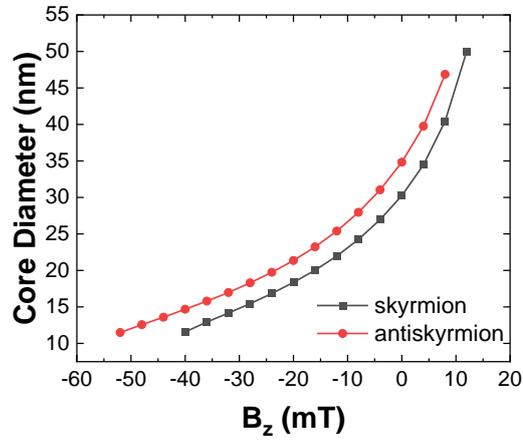

**Fig. 4:** Skyrmion and antiskyrmion core diameter under the perpendicular static magnetic field $B_z$.

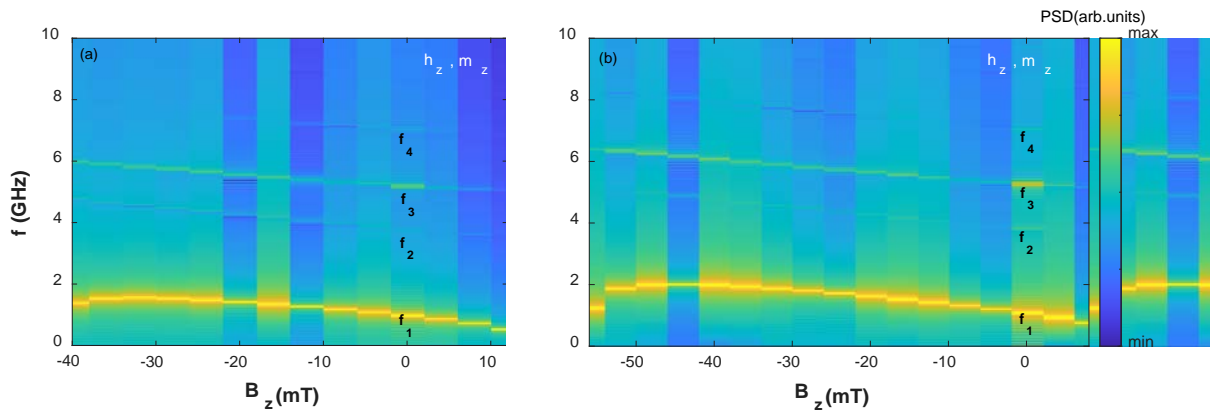

**Fig. 5:** Power spectral density color map of excitations for (a) skyrmion and (b) antiskyrmion under the static field $B_z$.

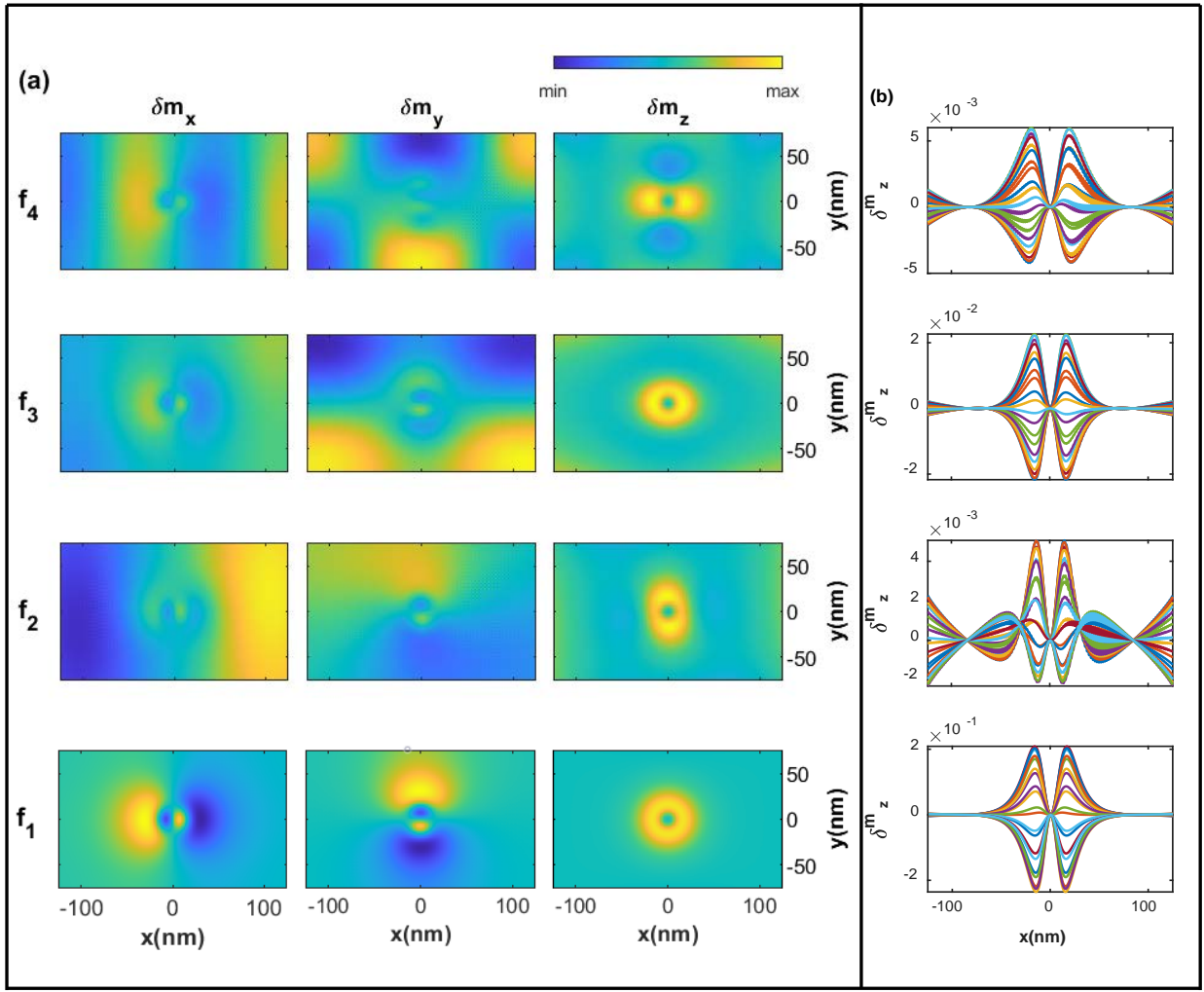

**Fig. 6:** Spatial fluctuations in the magnetization of skyrmion spin texture for $B_z = 0$ mT.
(a) Magnetization components spatial fluctuations for four modes ($f_1 - f_4$).
(b) Variation of $\delta m_z$ along the x - axis for one period of oscillation.

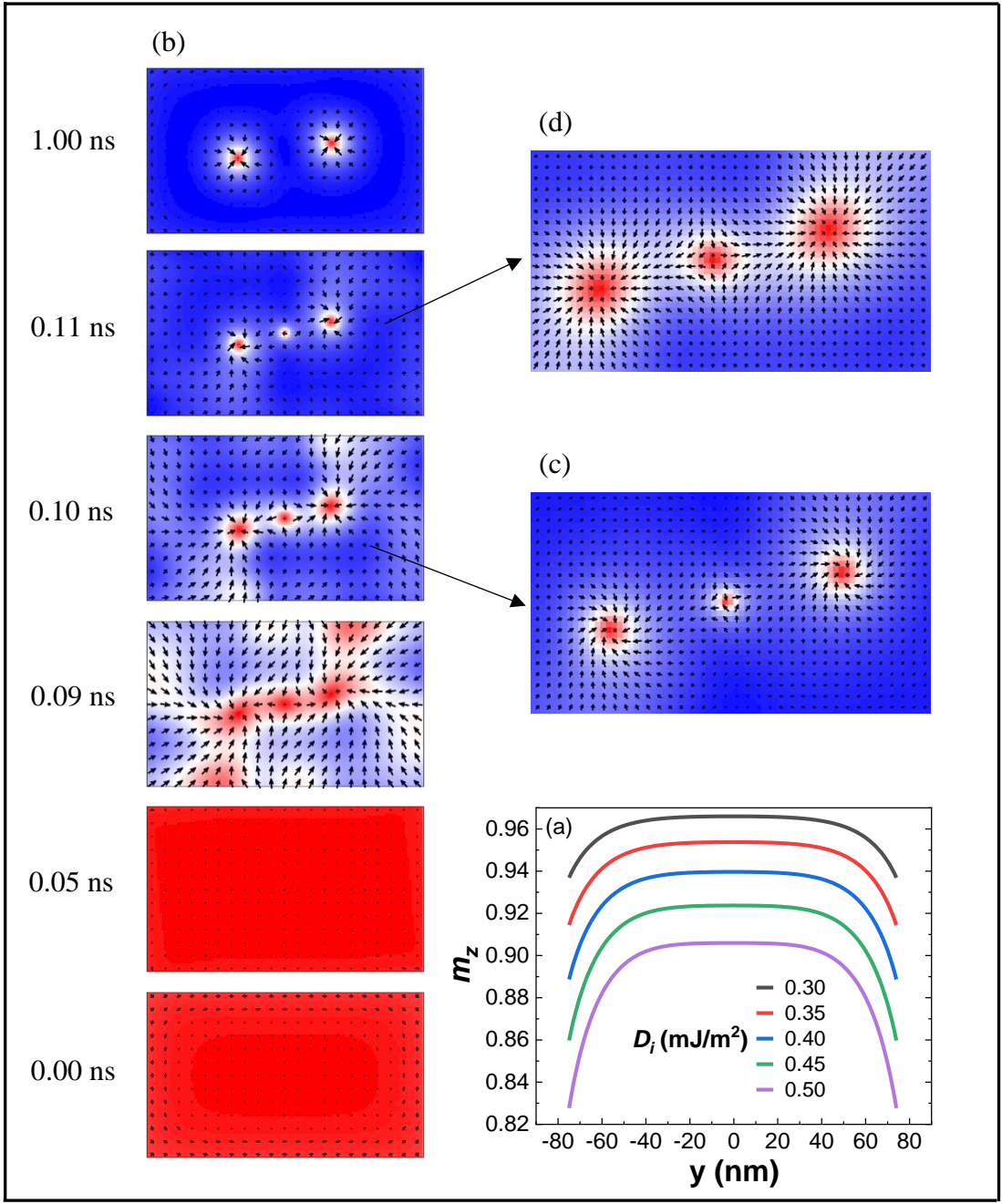

**Fig. 7:** (a) The magnetization component $m_z$ at the boundary edges along the width of the nanostructure for the different DMI values. (b) Spin states of skyrmion pair creation with an asymmetric current pulse for $D_i = 0.35$ mJ/m$^2$: $J_z = 3.0 \times 10^{12}$ A/m$^2$ applied for 50 ps ($t_p$) and $J_z = -3.0 \times 10^{12}$ A/m$^2$ applied for 70 ps ($t_n$). (c) and (d) are zooms of the spin states at 0.10 and 0.11 ns.

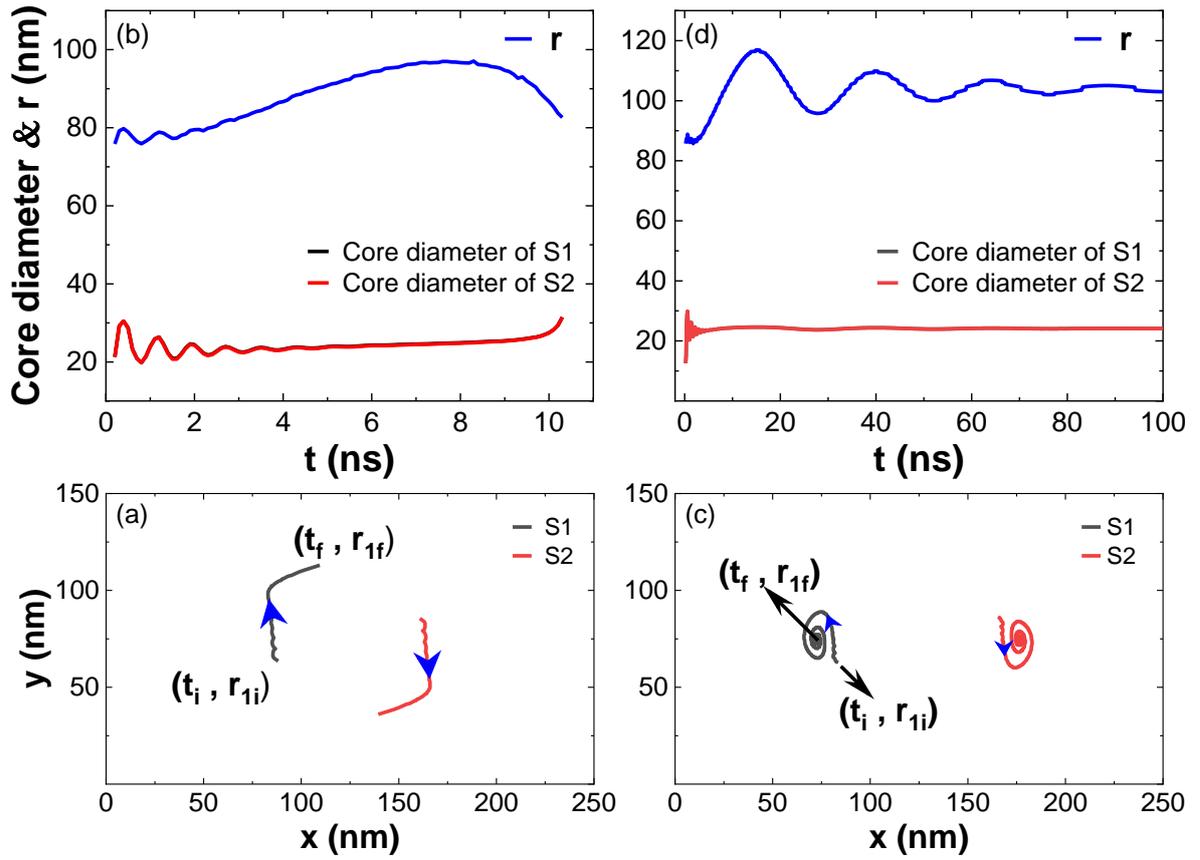

**Fig. 8:** Skyrmion pair dynamics in the relaxation for the different positive current pulse widths. (a) and (b) $t_p = 10$ ps. (c) and (d) $t_p = 50$ ps. S1: Skyrmion 1 and S2: Skyrmion 2. **r** is the separation distance between the skyrmions S1 and S2. $r_{1i}$ and $r_{1f}$, indicate the initial and final positions of skyrmion 1 (S1).

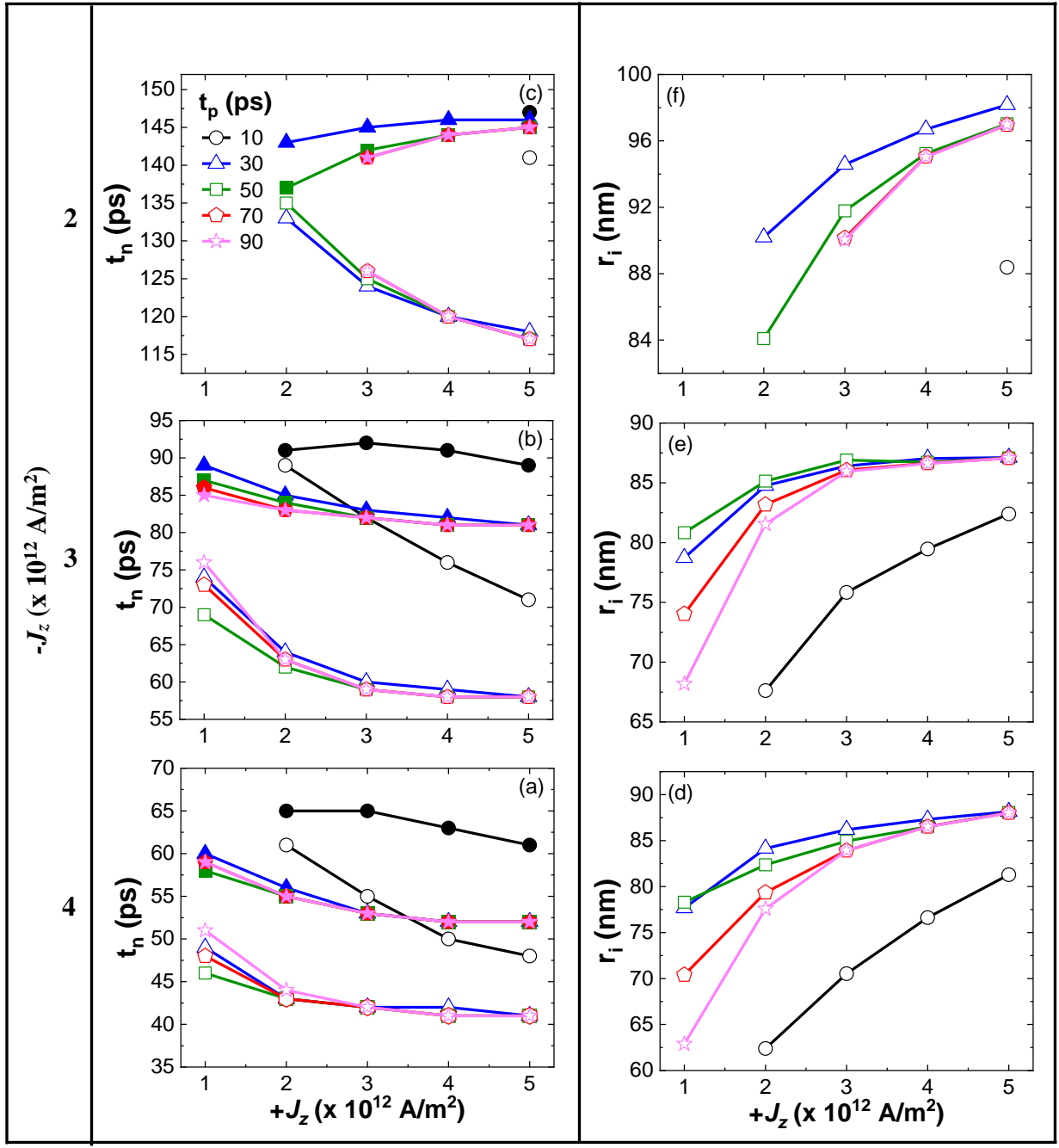

**Fig. 9:** (a), (b) and (c) show the negative pulse time ($t_n$) dependence on the positive and negative current densities for different positive pulse width ($t_p$) for the skyrmion pair formation. Hollow points indicate minimum $t_n$ and solid points indicate maximum $t_n$. (d), (e) and (f) show the initial separation between the skyrmion pair dependence on the positive current density ($+J_z$) and its pulse time ($t_n$). The initial separation is same for the different negative pulse times ($t_n$) under the fixed positive and negative current density and positive pulse time ($t_p$).

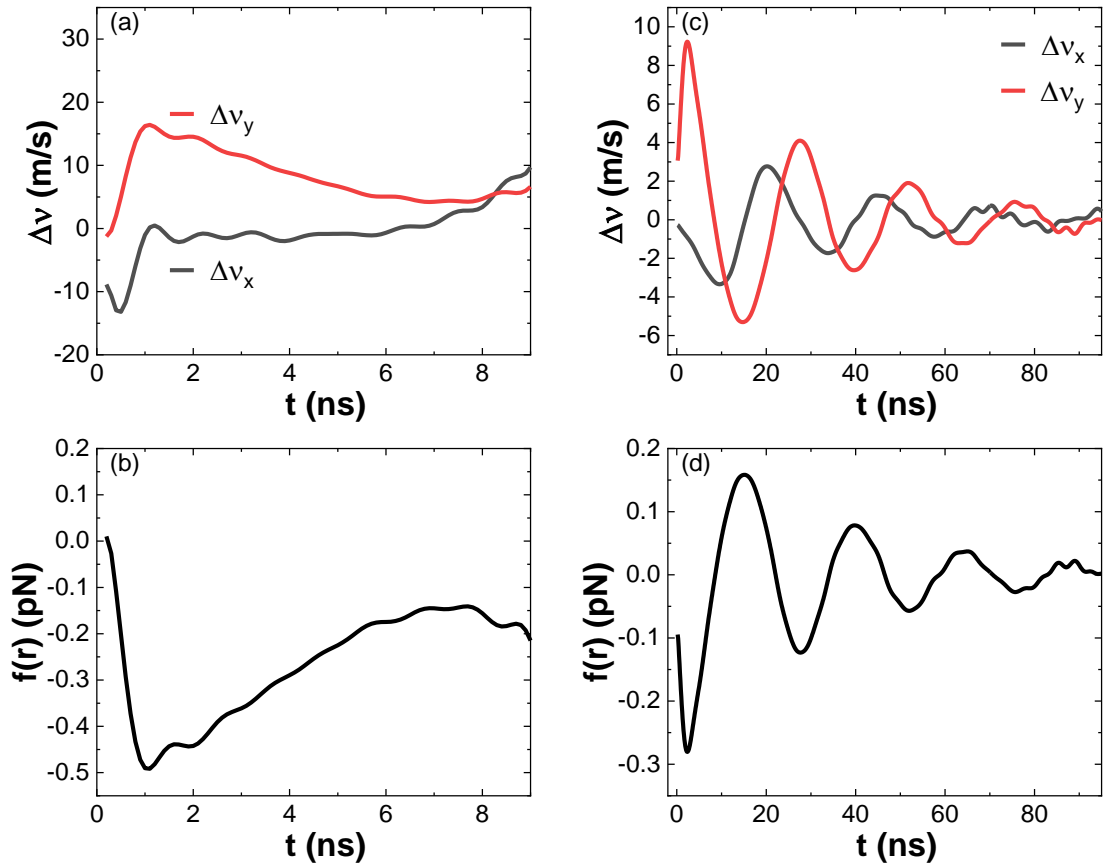

**Fig. 10:** The difference in velocities and interaction force variation for the unstable and stable skyrmion pair. (a) and (b) unstable skyrmion pair, (c) and (d) stable skyrmion pair (for the dynamics explained in Fig. 8).

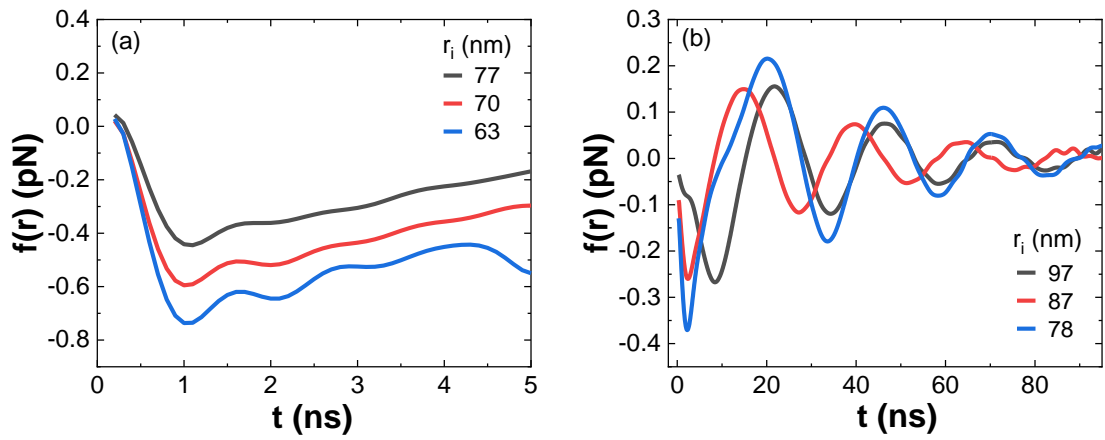

**Fig. 11:** Interaction force variation for the different initial distances between the skyrmion pair. (a) Unstable skyrmion pair. (b) Stable skyrmion pair.

# Supplementary material: Chiral spin textures creation and dynamics in a rectangular nanostructure

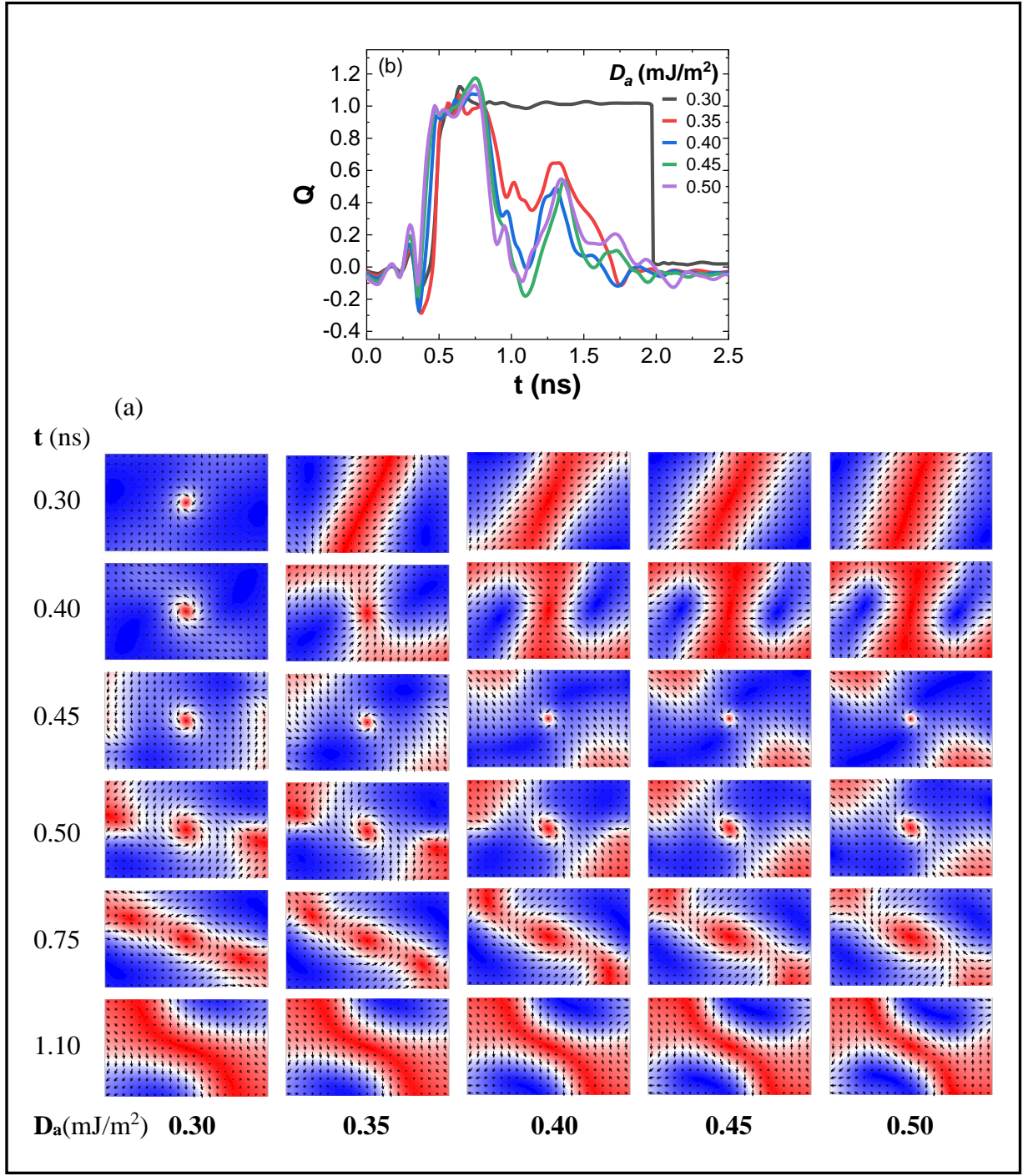

**Fig. S1:** Spin states and corresponding topological charge variation for $J_z = -0.5 \times 10^{12}$ A/m$^2$. (a) The spin states at different instants for $D_a = 0.30 - 0.50$ mJ//m$^2$ show the evolution of skyrmion. (b) The topological charge Q = 1 for the corresponding spin states indicates that the spin texture is a skyrmion.

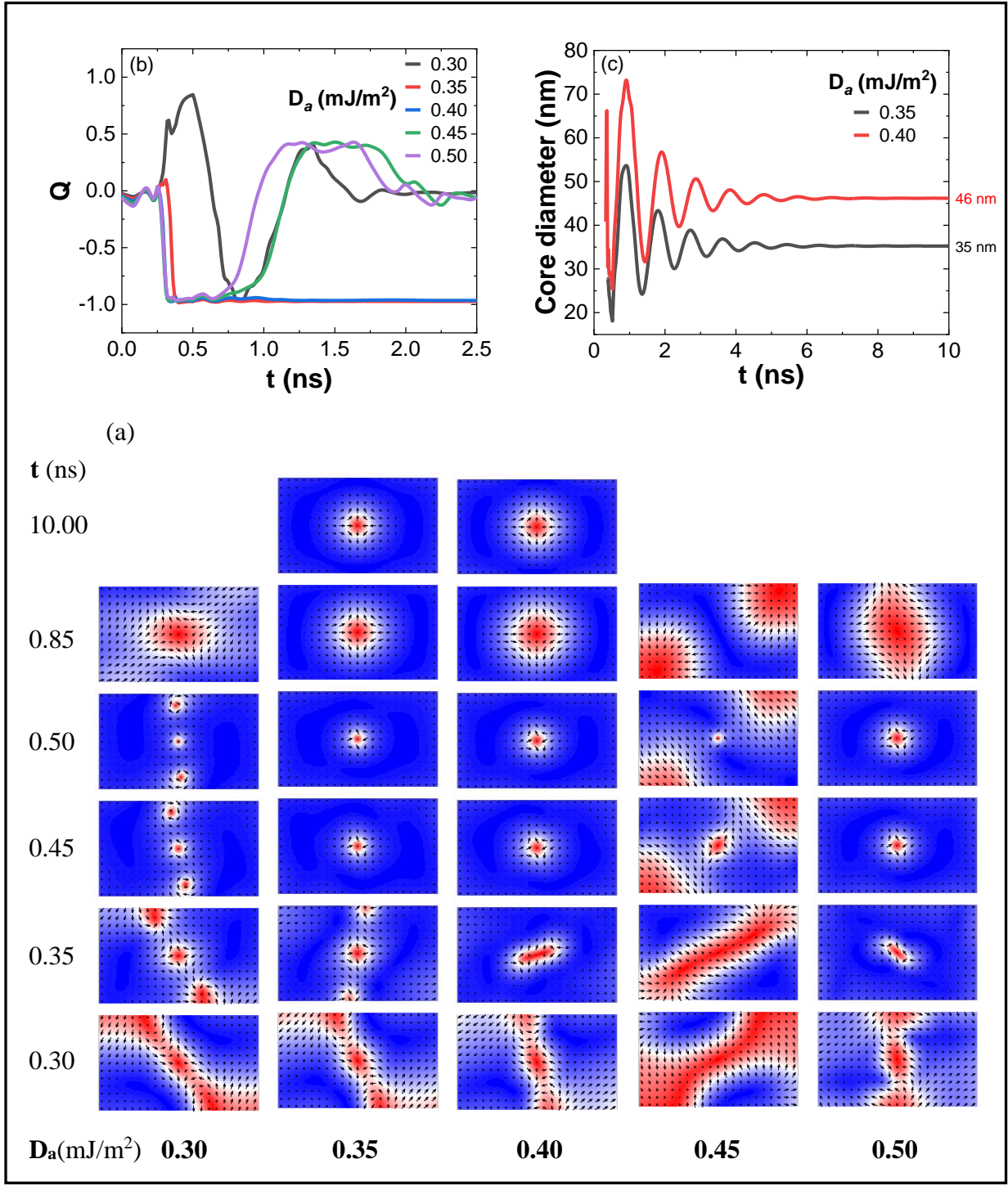

**Fig. S2:** Spin states, topological charge Q and the antiskyrmion core diameter variation for $J_z$ = -0.6 x $10^{12}$ A/m$^2$. (a) Evolution of stripe-like domain is ripped off into an antiskyrmion and two incomplete skyrmions for $D_a$ = 0.30 and 0.35 mJ/m$^2$. Skyrmion is stabilized for $D_a$ = 0.35 – 0.40 mJ//m$^2$. (b) The topological charge Q = -1 at 10 ns indicates a stable antiskyrmion. (c) The core diameter variation up to ~ 5 ns indicates the breathing mode, and the antiskyrmion size increases with DMI strength.